\documentclass[apj]{emulateapj}
\usepackage{graphicx,url,amssymb,amsmath,rotating,units,wasysym,listings,bm}
\usepackage{natbib,verbatim,enumerate}

\def\Me{M_{\rm ext}}
\def\Ren{R_{\rm ext}}
\def\Ee{E_{\rm ext}}
\def\ve{v_{\rm ext}}
\def\Rc{R_{\rm core}}
\def\Mc{M_{\rm core}}
\def\Ms{M_\odot}
\def\E51{E_{51}}

\newcommand{\lp}{\left(}
\newcommand{\rp}{\right)}

\newcommand{\ee}{\end{eqnarray}}
\newcommand{\be}{\begin{eqnarray}}

\begin{document}
\normalsize

\slugcomment{Accepted for publication in The Astrophysical Journal}
\shorttitle{Double-peaked Supernovae}
\shortauthors{Nakar \& Piro}

\author{Ehud Nakar\altaffilmark{1} and Anthony L. Piro\altaffilmark{2}}

\altaffiltext{1}{Raymond and Beverly Sackler School of Physics and Astronomy, Tel Aviv University, Tel Aviv 69978, Israel}

\altaffiltext{2}{Theoretical Astrophysics, California Institute of Technology, 1200 E California Blvd., M/C 350-17, Pasadena, CA 91125, USA}

\title{Supernovae with two peaks in the optical light curve and the signature of progenitors with low-mass extended envelopes}

\begin{abstract}
Early observations of supernova light curves are powerful tools for shedding light on the pre-explosion structures of their progenitors and their mass-loss histories just prior to explosion. Some core-collapse supernovae that are detected during the first days after the explosion prominently show two peaks in the optical bands,  including the $R$ and $I$ bands, where the first peak appears to be powered by the cooling of shocked surface material and the second peak is clearly powered by radioactive decay. Such light curves have been explored in detail theoretically for SN 1993J and 2011dh, where it was found that they may be explained by progenitors with extended, low-mass envelopes. Here we generalize these results. We explore, first, whether any double-peaked light curve of this type can be generated by a progenitor with a ``standard'' density profile, such as a red supergiant or a Wolf-Rayet star. We show that a standard progenitor (1) {\it cannot} produce a double-peaked light curve in the $R$ and $I$ bands, and (2) {\it cannot} exhibit a fast drop in the bolometric luminosity as is seen after the first peak. We then explore the signature of a progenitor with a compact core surrounded by extended, low-mass material. This  may be a hydrostatic low-mass envelope or material ejected just prior to the explosion. We show that it naturally produces both of these features. We use this result to provide simple formulae to estimate (1) the mass of the extended material from the time of the first peak, (2) the extended material radius from the luminosity of the first peak, and  (3) an upper limit on the core radius from the luminosity minimum between the two peaks. 
\end{abstract}

\keywords{ supernovae: general ---
supernovae: individual (SN 1993J, SN 2011dh, SN 2006aj)}

\section{Introduction}
The first hours to days of a supernova (SN) light  curve holds  valuable information on the structure of the  progenitor and on its  mass-loss history before the  explosion. However,  until recently, only a small  number of events were caught sufficiently early to extract this information. This has changed with the advent of sensitive, large field  of  view,  transient surveys, such as the Katzman Automatic Imaging Telescope \citep[KAIT;][]{Filippenko01}, the Palomar Transient Factory \citep[PTF;][]{Rau09} and the Panoramic Survey Telescope and Rapid Response System \citep[Pan-STARRS;][]{Kaiser02}. In the future, these efforts will continue to grow with SkyMapper \citep{SkyMapper}, the Zwicky Transient Facility \citep[ZTF;][]{Law09}, the All-Sky Automated Survey for Supernovae \citep[ASAS-SN;][]{Shappee14}, and the Large Synoptic Survey Telescope \citep{LSST09}. Today a growing number of SNe  are detected within one or two days of the explosion, opening  a new window into the relatively unexplored early phase of these events.

\begin{figure}
\epsscale{1.2}
\plotone{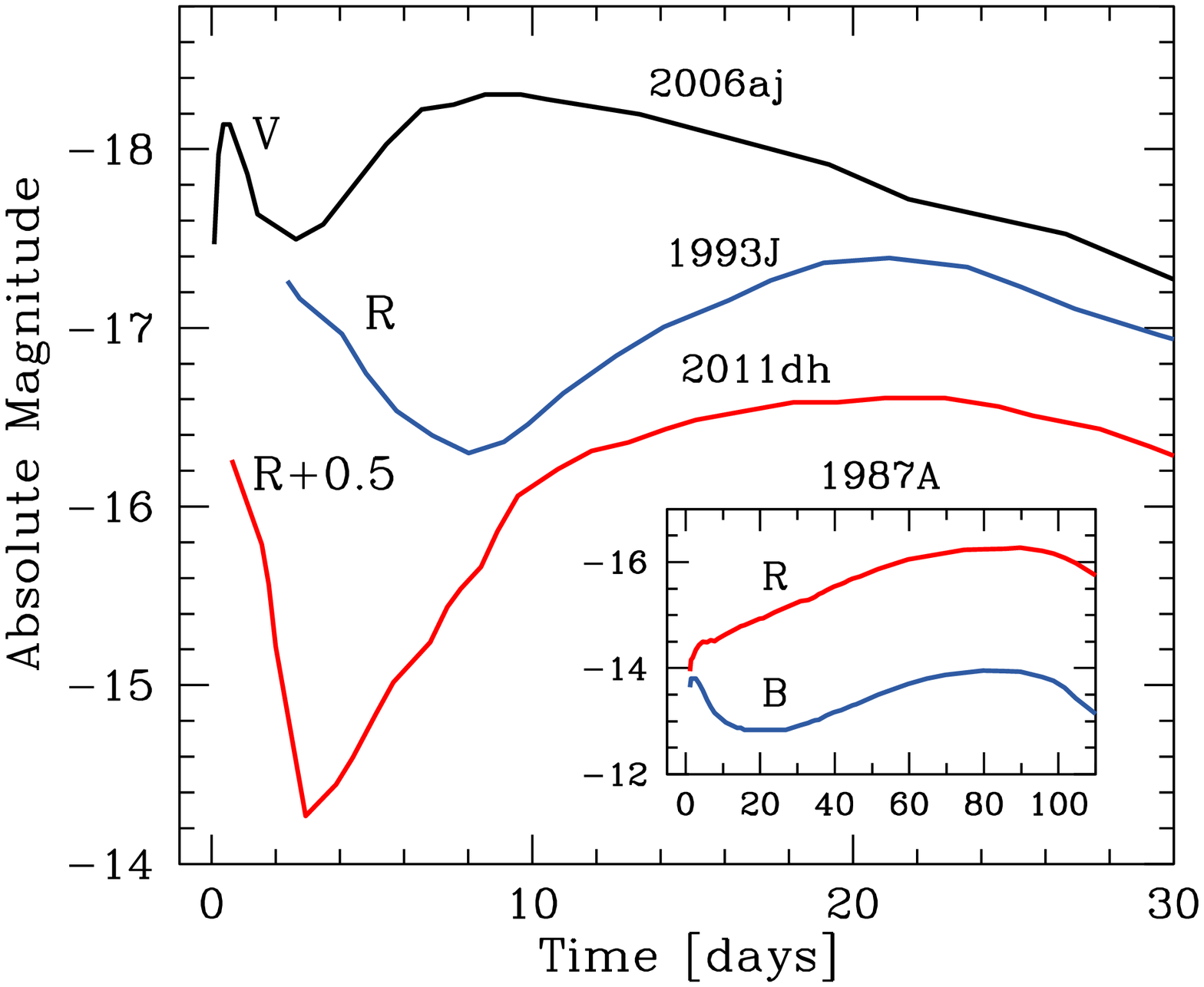}
\caption{The observed $V$ band light curve of the broad-line Ic SN 2006aj \citep{Campana06} and the $R$ band light curves of the Type IIb SNe 1993J \citep{Richmond94} and 2011dh \citep{Arcavi11}. These are all examples of observed light curves with two peaks of the type that we consider here. Namely, a second peak that is clearly powered by $^{56}$Ni and a first peak observed in red optical bands. SNe with such light curves cannot have standard progenitors, where a large fraction of the progenitor mass reaches out to the stellar radius, and are most likely generated by progenitors with a compact core surrounded by an extended low-mass material. The inset shows the $R$ and $B$ band light curves of SN 1987A \citep{Hamuy88}, where two peaks are observed only in $B$ while in redder bands ($V$, $R$ and $I$) only a single peak is observed. In that case the progenitor was a standard blue supergiant.}
\label{Fig1}
\epsscale{1.0}
\end{figure}

Following the  detection of some very young SNe,  an unexpected discovery has been that in a subset of  these SNe the  optical light  curve, including  $R$ and  $I$  bands, show two prominent  peaks. In these  events the second optical peak  is on a time scale  of weeks and is clearly  powered by the
decay of $^{56}$Ni, while  the first peak fades on a time  scale of days. The
best  known  example  of  such  a  light  curve  is  the  Type  IIb  SN  1993J
\citep{Wheeler93}. More  recent examples are  the Type IIb SNe  2011dh (a.k.a.
PTF11eon;  \citealt{Arcavi11}) and  2013df \citep{VanDyk13},  the Type  Ibn SN
iPTFbeo  \citep{Gorbikov13},  and   the  Type  Ic  (broad-lined)  SN  2006aj
\citep{Campana06}. Thus, many of these  events are of Type IIb, but they also include core-collapse SNe of other types. Several
examples of observed light curves are depicted in Figure \ref{Fig1} (SNe  2006aj, 1993J and 2011dh). This shows  how the first peak can rival or
exceed  the luminosity  of  the  second peak  as  well  as the  characteristic
time scale  of each  peak.  In all  these  cases the  first  peak is  observed
simultaneously in the red and the blue bands. This is different than the typical case where only a single peak is observed in the red bands, even if two peaks are observed in blue and UV light (e.g., SN 1987A, see the inset of Figure \ref{Fig1}).

SNe 1993J,  2011dh and 2013df  are among the  rare cases where progenitors were  identified in  pre-explosion images.  All three of these SNe were found to be supergiants with radii $\gtrsim 10^{13}\,{\rm cm}$ \citep{Aldering94,Maund11,VanDyk11,VanDyk13}. Such a large radius was claimed to be in tension with the absence of bright, long-lived emission, as would have been expected from  a cooling of shocked extended envelope \citep{Arcavi11}. This discrepancy was explained by invoking a low mass for the envelope \citep{Hoflich93,Woosley94,Bersten12}.

Motivated by these discoveries and previous theoretical work on SNe 1993J and 2011dh, we investigate the conditions required to produce SN light curves with two peaks of this type, i.e., second peak powered by radioactive decay and first peak observed in all optical bands (including the red bands) on time scales of hours to days. We then summarize what can be learned from such observations. We divide our discussion between ``standard''  core-collapse progenitors, where a large fraction of the mass reaches out to the stellar radius, and ``non-standard'' progenitors, which have a compact core surrounded by extended, low-mass material. In \S \ref{Sec:CoolingEnvelopeLimits}, we show that the standard progenitors cannot produce an early peak in the $R$ and $I$ bands nor in their bolometric light curves (two prominent peaks may be still seen only the blue optical bands and the UV).   Interestingly, standard progenitors with extended envelopes (e.g., red supergiants) are predicted to  produce a peak  in all  bands  $\sim10\,{\rm min}$ after the  shock breakout, followed by  a slow decay over several hours \citep{Nakar10}.  This  feature has yet to be detected, but its discovery would be an important test for the understanding of these massive stars.

In \S \ref{Sec:ExtendedEnvSignautre}, we show that the density profiles of non-standard progenitors naturally produce two peaks observed in all optical bands, with a sharp drop of the bolometric luminosity between the peaks, on a time scale of hours to days after the explosion. We then provide simple relations to be used in conjunction with observations of these events to constrain the mass (eq.~[\ref{eq:Menv}]) and radius (eq.~[\ref{eq:Ren}]) of  the low-mass extended material, along with  the  radius  of the core (eq.~[\ref{eq:Rcor}]). These are confirmed with comparisons to previous detailed modeling. We conclude with a summary of our results in \S \ref{Sec:Summary}.

\section{Standard Progenitors}\label{Sec:CoolingEnvelopeLimits}

In this section we explore the expected light curve from standard core-collapse SN progenitors, in which most of the mass is  
concentrated near the stellar radius, $R_*$. To understand what is meant by this, consider two typical cases. The first is an extended progenitor, such as a red supergiant, which has a massive hydrogen envelope in hydrostatic equilibrium. Here, $\Me > \Mc $ and $\Ren = R_* \gg \Rc$, where $\Mc$ and $\Rc$ are the core  ejected mass\footnote{This is the mass from the top of the helium core inward, minus the remnant mass ($\approx1.4\,M_\odot$) left over that will produce a neutron star.} and radius, respectively, and $\Me$ and $\Ren$ are the  mass and radius of the extended envelope, respectively (a more specific definition of $\Me$ in the context of this paper is given later). The second is a stripped progenitor, such as a Wolf-Rayet (WR) star. Although such stars have little or no envelope, most of the mass is again concentrated near  $R_*$. In hydrostatic equilibrium the density profile, $\rho(r)$, at a radius $r \approx R_*$ varies on a scale that is comparable to the distance from the stellar edge. We approximate it by a polytrope $\rho \propto x^n$, where $x=(R_*-r)/R_*$ and $n$ is typically in the range $1-3$. This approximation is expected to be good for a red supergiant although it may be a bit simplistic for a WR, where radiation close to the Eddington luminosity may affect the density profile near the edge.

Near the stellar edge, $x \ll 1$, the SN shock accelerates with the decreasing density as $v \propto \rho^{-\beta}$, where the value of $\beta$ depends weakly on $n$ \citep{Sakurai60,Grassberg81}. For standard progenitors with $n=1-3$, $\beta = 0.19$ while for $n=15$, an extremely steep density profile, $\beta =0.17$. Hereafter we use $\beta=0.19$. The shock heats and accelerates the material and after it breaks out of the stellar edge, the observed luminosity is determined by the diffusion of photons through the hot expanding gas. The light curve of this cooling phase for progenitors with a $\rho\propto x^n$ density profile has been calculated analytically by many authors \citep[e.g.,][]{Chevalier92,Piro10,Nakar10,Rabinak11}. Here we focus on the results from \cite{Nakar10}, which calculated the observed temperature most accurately.

\subsection{Planar Phase}\label{Sec:Planar}

At first, before the gas roughly doubles its radius, the evolution of the surface layers is planar (as is discussed in more detail by \citealt{Piro10,Nakar10}). Here we highlight the main results of the optical emission during this phase. A detectable optical emission is only expected if the shock breakout radiation is in thermal equilibrium, namely if the progenitor is a supergiant with an extended envelope. In a supergiant progenitor, the optical light curve peaks on a time scale of $R_*/c$ after the explosion, while the planar phase lasts for a time $R_*/v$. For a progenitor with $R_*\approx500\,R_\odot$, it implies a peak after $\approx1000\,{\rm s}$ and an end to the planar phase at $\approx10\,{\rm hr}$. In compact progenitors, the optical emission during the breakout and the planar phase is too faint to be observed with current instruments. 

At any given time the observed luminosity is generated at a mass depth $m_{\rm obs}$ (measured from the outside inward), where the diffusion time equals the dynamical time. During the entire planar phase photons diffuse out from the breakout layer (i.e., $m_{\rm obs}$ is roughly constant and equal to the mass from where the shock breaks out; see \citealt{Nakar10} for details). The resulting light curve evolves as
\begin{equation}\label{eq:planarLT}
 \begin{array}{ccc}
  L_{\rm bol} & \propto & t^{-4/3}, \\
 &&  \\
  T_{\rm obs} &\propto & t^{-0.35}.
 \end{array}
\end{equation}
Since $T_{\rm obs}$ is in the UV, the optical flux scales as  $F_{\nu,\rm opt}\propto L/T_{\rm obs}^3$, and drops during this phase at a slow rate of
\begin{equation}\label{eq:Fnu}
 F_{\nu,\rm opt}\propto t^{-0.28} .
\end{equation}
Thus, during the entire planar phase the optical flux is expected to decrease by $\approx1\,{\rm mag}$. During this phase all the optical bands are on the Rayleigh-Jeans tail.

To conclude, in the case of a standard extended hydrostatic envelope the first observed emission is an optical/UV peak with a very short rise time of minutes, followed by a much slower decay of $\approx1\,{\rm mag}$ over the next several hours. The emission then starts rising again as the spherical phase begins.

\subsection{Spherical Phase}
After the gas roughly doubles its radius, the spherical phase of the expansion begins. Now the depth in mass from which photons diffuse out increases rapidly with time, such that at a time $t$ the observed mass is
\be\label{eq:mobs}
	m_{\rm obs}=5 \times 10^{-3} \kappa_{0.34}^{-1}v_9
	\left( \frac{t}{1\,{\rm day}}\right)^2
	M_\odot,
\ee
where $\kappa$ is the opacity with $\kappa_{0.34}=\kappa/0.34\,{\rm cm^2\,g^{-1}}$, and $v$ is the velocity\footnote{Equation \ref{eq:mobs} is implicit since $v$ is in itself  a function of $m_{\rm obs}$. However, later we will discuss methods to estimate $v$ independently for the mass of interest.} of $m_{\rm obs}$ with $v_9=v/10^9\,{\rm cm\,s^{-1}}$.

During the spherical phase, the bolometric luminosity drops as a power-law with
\be\label{eq:Lobs}
	L_{\rm bol} \propto t^{-\alpha},
\ee
 where \citep{Nakar10}
\be\label{eq:alpha}
	\alpha={\frac{2.28n-2}{3(1.19n+1)} } < 0.64.
\ee
The upper limit for $\alpha$ is derived for very large values of $n$, i.e., an unrealistically sharp drop in the density. Therefore equation (\ref{eq:alpha}) shows that there is a limit to how quickly the bolometric luminosity can fall. More realistically, for the canonical values of $n=1.5$ (convective envelope) and $n=3$ (radiative envelope), $\alpha=0.17$ and $\alpha=0.35$, respectively. Thus, for any standard progenitor the bolometric luminosity decrease during this phase can be at most moderate. {\it If a more rapid luminosity drop is observed, it implies that either the density structure is highly non-standard or that the diffusion front has travelled through the entire envelope} (i.e., $m_{\rm obs}>\Me$). The latter, for example, is the origin of the fast drop seen from Type II-P SNe at the end of their plateau phase.

Another limit on double-peaked light curves can be derived with respect to the $R$ and $I$ band properties. This can be seen because during the spherical phase, before recombination becomes important, the temperature evolves roughly  as
\be\label{eq:Tobs}
	T_{\rm obs} \propto t^{-0.6},
\ee
where the dependance on $n$ is weak. As a result, the observed flux in bands that are on the Rayleigh-Jeans tail of the spectrum rises as $t^{1.5}$ \citep{Piro13}. The flux starts falling only once the temperature falls to the point that the observed band is on the Wein part of the spectrum. However, once the observed temperature reaches about $6000-8000\,{\rm K}$ the ionization level of the gas drops significantly. This has two effects. The most prominent one is that the observed temperature drop stops almost entirely. Thus, the roughly constant temperature is set so the peak of observed spectrum is around the $R$ and $I$ bands. The second is that the bolometric luminosity falls more slowly (or even start rising slowly) when the recombination front reaches deep enough to affect $m_{\rm obs}$. The result is that as long as $m_{\rm obs}<\Me$ the $R$ and $I$ band luminosities are rising while the temperature is higher than about $6000-8000\,{\rm K}$ and it 
remains rather constant after it drop to this level (this is the origin of the plateau in Type II-P SNe). This is in contrast to the optical blue bands and UV, which are at the Wein part of the observed spectrum at a temperature of $6000-8000\,{\rm K}$. Thus, even the mild decrease in the observed temperature results in a significant drop of the blue light. This result is true both for hydrogen rich envelopes and for hydrogen striped progenitors \citep{Dessart11}.

To conclude, before the gas ionization level drops, the $R$ and $I$ bands are rising. After it drops, these bands are rather constant, or at most the $R$ band is dropping very slowly. This implies that the cooling envelope phase of a standard progenitor with a massive envelope cannot produce a prominent peak in the $R$ or $I$ band as long as $m_{\rm obs}<\Me$.

\section{Non-standard Progenitors}\label{Sec:ExtendedEnvSignautre}

\begin{figure}
\epsscale{1.1}
\plotone{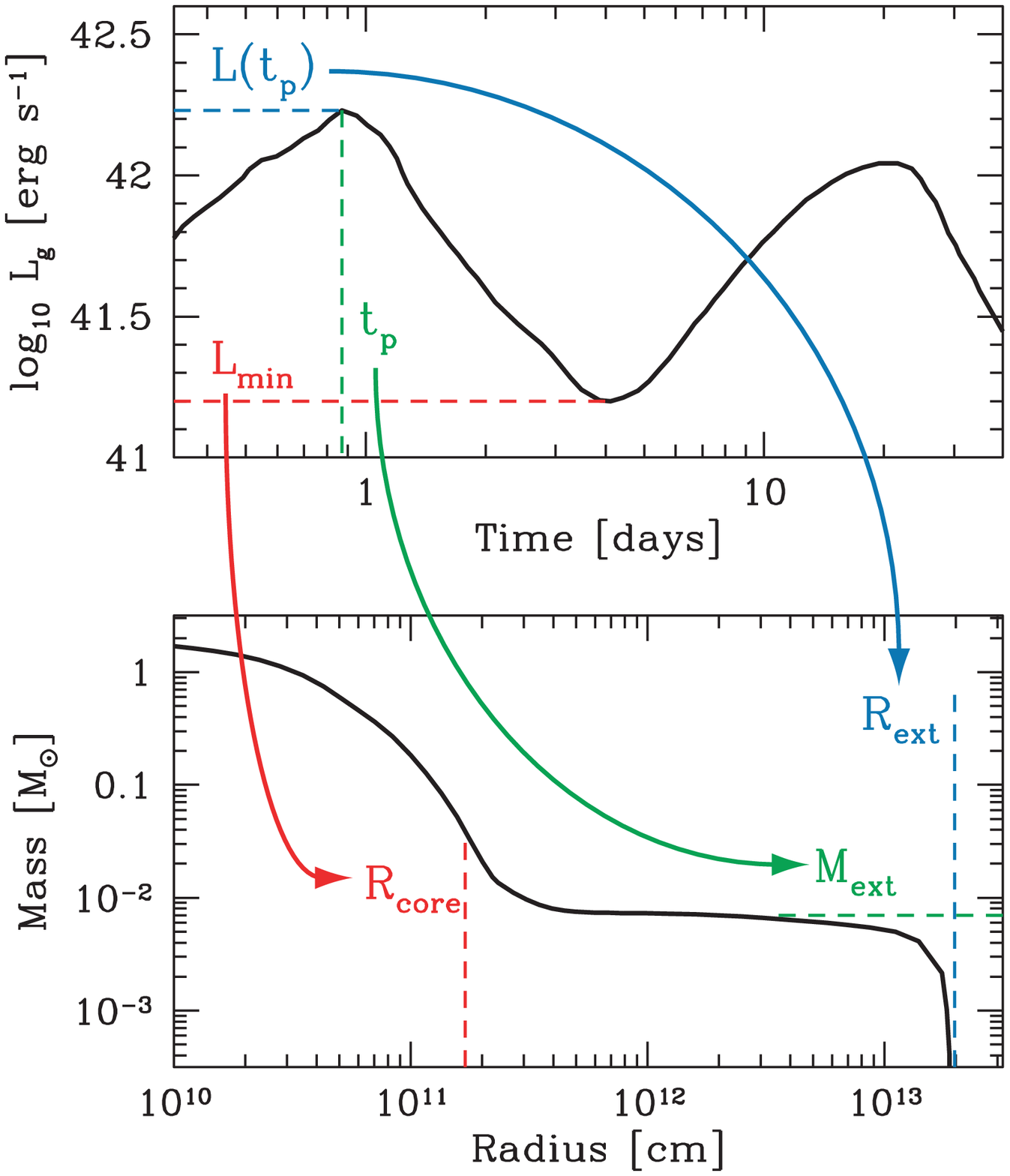}
\caption{The top panel shows an example a double-peaked SN optical light curve calculated numerically by \citet{Bersten12}. The light curve shape is similar in all optical bands (here we present $g'$-band luminosity). The bottom panel shows the structure of a non-standard progenitor used for the calculation of the light curve. Mass is measured from the stellar edge inward. Arrows and color-coding highlight which aspects of the SN light curve provide information about the progenitor structure. The luminosity of the first peak $L(t_p)$ provides an estimate of the stellar radius $\Ren$ (shown in blue). The mass of the envelope $\Me$ (roughly $6 \times 10^{-3}\,M_\odot$ as taken from $\Ren/3$ to $\Ren$) is estimated by the time of the first peak $t_p$ (shown in green). The minimum luminosity $L_{\rm min}$, provides an upper limit on the core radius $\Rc$ (shown in red).}
\label{Fig2}
\epsscale{1.0}
\end{figure}

Motivated by the inability of standard progenitors to reproduce the main features of double-peaked light curves of the type we consider here, we now turn to considering non-standard progenitors. In particular, since the standard progenitors appear to fail when $m_{\rm obs}<\Me$ we look at lower amounts of material surrounding a compact core\footnote{Given that typical cores are expected to be with $\Rc \sim R_\odot$ and $\Mc$ of about several $\Ms$ or more, the typical extended material radius that we consider are $\Ren \gtrsim 10^{12}$ cm and its mass is $\Me \ll \Ms$.}, i.e., $\Me \ll \Mc$ and $\Ren \gg \Rc$. In such cases, $m_{\rm obs}<\Me$ will not be satisfied for long during the light curve evolution. An example of a non-standard progenitor is shown in the lower panel of Figure \ref{Fig2}. We plot the mass measured from the stellar edge inward to highlight just how little mass is in the extended material. In this example, $M_{\rm ext}\approx6\times10^{-3}\,M_\odot$, even though it constitutes the outer $2/3$ of the star in radius!

The exact density profile of the extended material is unimportant for our analysis. The only important properties are that $\Me$ is concentrated around $\Ren$ and that the density at radii larger than $\Ren$ is low enough so interaction can be neglected. Thus, the extended material can be a shell ejected just prior to the explosion \citep[e.g.,][]{Ofek13} or a continuous wind, as long as it is terminated at $\Ren$. It can also be a low-mass extended envelope, either in or out off hydrostatic equilibrium. Note that in that case the mass $\Me$ is not strictly the envelope mass. The reason is that $\Me$ includes only mass that is concentrated around $\Ren$, while some envelope mass may be found at smaller radii (when we look at the profiles of specific models later in this section, it will be more clear why we must make this distinction). We also restrict the discussion here to cases where
\be\label{eq:me}
	\Me\gtrsim\frac{4\pi\Ren^2}{\kappa}\frac{c}{v}
	= 5 \times10^{-5} \kappa_{0.34}^{-1}v_9^{-1}R_{13}^2
	\,M_\odot,
\ee
where $R_{13}=\Ren/10^{13}\,{\rm cm}$. This criterion ensures that the shock breaks out from the extended material and not from the core\footnote{When the envelope mass is smaller than this criterion the shock breaks out of the core and the interaction of the expanding core with the envelope is seen directly by the observer. Such cases are out of the scope of this paper.}.

When there is low-mass, extended material, the physical picture changes as follows. After crossing the core, the shock accelerates the low-density material to rather high velocities. Adiabatic losses due to expansion of the shocked extended material are relatively small (due to its initial large volume) so its cooling emission is bright, dominating the early-time light curve. However, this emission falls off very rapidly once $m_{\rm obs}>\Me$, implying that if the extended material mass is low, then this phase ends within hours to  days (using eq. [\ref{eq:mobs}]). At that point the main source of the emission becomes the core. Here, adiabatic loses are severe as the radius before the expansion is much smaller, so that the main source of emission is the radioactive decay of $^{56}$Ni. The observed radioactive luminosity increases as more mass of the core, and thus of $^{56}$Ni, is exposed by the inward traveling diffusion front. The peak of this phase is observed roughly when $m_{\rm obs}\approx\Mc$ (note that $\Mc$ includes only the ejected core mass and not any potential remnant mass that is left over from the SN).

Therefore, low-mass, extended material around a compact core naturally leads to a double-peak SN light curve in all wavelengths, including the $R$ and $I$ bands. It also results in a sharp drop in the bolometric luminosity, between the end of the cooling phase and the emergence of the $^{56}$Ni driven core luminosity. Calculating the main properties of the resulting light curve is simplified by the fact that the emission of the extended material and the core are independent of each other. {\it One can in effect treat the emission as that of two separate SNe.} The first SN is the cooling phase emission of a low energy explosion of an extended low-mass star. This emission is short lived and the time, luminosity and temperature at the peak are straightforward to calculate since, as we show below, recombination does not play a role. The second SN is a regular compact star explosion which was calculated by many authors in the context of Ib and Ic SNe.  Below we discuss ways to estimate the extended material velocity and energy, and then we use these values to constrain the properties of the progenitor. Using arrow and color-coding, we highlight the connections between the double-peaked light curve and the progenitor structure in Figure \ref{Fig2}.

\subsection{Estimating the Velocity and Energy of the Extended Material}

The characteristic velocity of the extended material $\ve$ can be estimated from observations if an early spectrum of the first peak emission is available and the photosphere velocity at this time can be measured. The extended material velocity is smaller than the photospheric velocity at peak by a factor of order unity. For an envelope in hydrostatic equilibrium, this factor is in the range $1.3-1.5$ \citep{Nakar10}. Alternatively, a photospheric velocity at a later time can be used, since for a layer with a given mass $\tau \propto t^{-2}$. At the time of the first peak, the optical depth of $\Me$ satisfies $\tau \approx c/\ve$. Therefore, the optical depth of $\Me$ drops to unity roughly at $t \approx t_p\sqrt{c/\ve}$, where $t_p$ is the time of the first peak. For typical parameters this is at $\sim5\times t_p$, which is usually during the rising of the second peak. Thus, measuring the photospheric velocity at that time provides a good estimate of $\ve$.

If an observational constraint is not available, then $\ve$ can be estimated based on theory. Following the core collapse, a shock is driven through the remaining parts of the core. It accelerates once it encounters the sharp density drop at the edge of the helium core, bringing smaller amounts of mass to higher and higher velocities. This leads to a velocity profile $v(m_c)$, where $m_c$ is the amount of core mass accelerated to a velocity $v$. Once the shock starts propagating into the shallower density profile of the extended material it decelerates again, leading to a reverse-forward shock structure. During deceleration the swept-up extended material mass is comparable to the core mass that crossed the reverse shock, implying that by the time that the entire extended material is shocked its velocity is $\ve \approx v(m_c=\Me)$. We approximate $v(m_c)$ by assuming that the core density profile is not significantly affected by the extended material, in which case the acceleration follows the self-similar solution of \cite{Sakurai60} with $n=3$,
\be\label{eq:v_env}
 	\ve \approx 1.5 \times 10^9 \E51^{0.5}
 	\left(\frac{\Mc}{3 \Ms}\right)^{-0.35}
 	\nonumber\\
 	\times
 	\left(\frac{\Me}{0.01\Ms}\right)^{-0.15} {\rm cm\,s^{-1}},
\ee
where $E$ is the total explosion energy and $E_{51}=E/10^{51}\,{\rm erg}$. The energy carried by the extended material is then
\be\label{eq:E_env}
 \Ee \approx 2 \times 10^{49}  \E51 \left(\frac{\Mc}{3\Ms}\right)^{-0.7}
 \left(\frac{\Me}{0.01\Ms}\right)^{0.7}{\rm erg}.
 \nonumber
 \\
\ee
Thus, the cooling phase is similar to a low mass, low energy SN of an extended progenitor, which produces a bright, short-lived signal.

\subsection{Constraints on the Extended Material and the Core Properties}

The peak optical flux is observed when $m_{\rm obs}\approx\Me$. Thus, the mass of the extended material can be measured simply by identifying the time of the first optical peak, $t_{p}$. This is done by using equation \eqref{eq:mobs} only, when $\ve$ is measured from the observations, or by using also equation \eqref{eq:v_env} when it is not,
\be\label{eq:Menv}
 	\Me & \approx & 5 \times 10^{-3} \kappa_{0.34}^{-1} \lp \frac{\ve}{10^9\,{\rm cm\,s^{-1}}}\rp
	\left( \frac{t_{p}}{1\,{\rm day}}\right)^2M_\odot
		\nonumber\\
	& \approx & 8 \times 10^{-3} \E51^{0.43} \kappa_{0.34}^{-0.87}
	\left(\frac{\Mc}{3\,\Ms}\right)^{-0.3} \left( \frac{t_{p}}{1\,{\rm day}}\right)^{1.75}M_\odot.
		\nonumber\\
\ee
The connection between $t_p$ and $M_{\rm ext}$ is shown in green in Figure \ref{Fig2}. As we discuss above the emission from the extended material is dominated by the mass at $r \approx \Ren$. Thus, $\Me$ measures only the mass concentrated  at $r \approx \Ren$. If the envelope structure is such that a significant amount of mass is concentrated at $r \ll \Ren$, then this mass does not contribute to the flux at $t_{p}$ and is therefore not included in $\Me$. Note that $t_{p}$ is also roughly the decay time scale of the observed flux after the peak. So even if the SN is detected only after the peak, then the decay time scale can provide a rough estimate of $\Me$.

The bolometric luminosity at the peak is set by the initial internal energy in the extended material and the adiabatic loses to expansion, namely
\be
	L_{\rm bol}(t_{p}) \sim \frac{\Ee \Ren}{\ve t_{p}^2}.
\ee
Thus, the peak
emission also provides an estimate of the extended material radius,
\be\label{eq:Ren}
	 \Ren & \approx & 2 \times 10^{13} \kappa_{0.34} L_{43}
	 \left(\frac{\ve}{10^9\,{\rm  cm\,s^{-1}}} \right)^{-2}{\rm cm},
	 	\nonumber\\
 	& \approx & 10^{13} \kappa_{0.34}^{0.74} \E51^{-0.87} L_{43}
 	\left(\frac{\Mc}{3\,\Ms}\right)^{0.61}
	\left( \frac{t_{p}}{1\,{\rm day}}\right)^{0.51} 
	{\rm cm},
		\nonumber\\
\ee
where $L_{43}=L_{\rm bol}(t_p)/10^{43}\,{\rm erg\,s^{-1}}$. The observed temperature at the peak can be approximated by the effective temperature, resulting in
\be\label{eq:Tpeak}
	T_{\rm obs}(t_p) \approx 3\times10^4
 	\kappa_{0.34}^{-1/4}\left( \frac{t_{p}}{1\,{\rm day}}\right)^{-1/2}
 	\left(\frac{\Ren}{10^{13}\,{\rm cm}}\right)^{1/4}{\rm~K}.
	\nonumber\\
\ee
This temperature justifies ignoring recombination. It peaks in the UV and therefore cannot be measured easily by optical surveys, although it  may be possible to probe using future UV surveys \citep[e.g.,][]{Sagiv13}. Equations (\ref{eq:Ren}) and (\ref{eq:Tpeak}) enable constraints to be placed on $\Ren$ with optical photometry alone. The connection between $L(t_p)$ and $\Ren$ is shown in blue in Figure~\ref{Fig2}. Note, however, that if $\Ren$ is derived in this way, then for an observed frequency in the Rayleigh-Jeans tail $\Ren \propto L_\nu^4$. This implies that the order of unity uncertainty in the coefficients of equations (\ref{eq:Ren}) and (\ref{eq:Tpeak}) translates to an uncertainty of an order of magnitude in the derived $\Ren$.

  \begin{deluxetable*}{ccccccccccc}
  \tablecolumns{10} \tablewidth{0pt}
  \tablecaption{Comparison to Numerical Results}
  \tablehead{
    \multicolumn{1}{c}{} & \multicolumn{7}{c}{Numerical values} & \multicolumn{3}{c}{Equations (\ref{eq:Menv}), (\ref{eq:Ren}), and (\ref{eq:Rcor})\tablenotemark{a}} \\
    & $t_{p}$\tablenotemark{b} & $L(t_p)~({\rm erg\,s^{-1}})$\tablenotemark{c} & $L_{\rm min}$ & $M_{\rm core}$ & $\Me$\tablenotemark{d} & $\Ren$ &
   	$R_{\rm core}$ & $\Me$ & $\Ren$ &  $R_{\rm core}$ \\
    Ref. & (days)  & $L_\nu(t_p)~(M_g)$ & $({\rm erg\,s^{-1}})$ & ($M_\odot$) & ($M_\odot$) & ($10^{13}\,{\rm cm}$) & 
	($10^{11}\,{\rm cm}$) & ($M_\odot$) & ($10^{13}\,{\rm cm}$) & ($10^{11}\,{\rm cm}$)
    }
  \startdata
  B12 &0.27 & $M_{g}=-15.5$ & $2\times10^{41}$ & 2.5 & $4\times10^{-4}$ &1.1 & 1.7 & $7\times10^{-4}$ & 2 & $<5$ \\
  B12 &0.5 & $M_{g}=-16.2$ & $2\times10^{41}$ & 2.5 & $2\times10^{-3}$ & 1.4 & 1.7 & $2\times10^{-3}$ & 2.4 & $<5$ \\
  B12 &0.85 & $M_{g}=-16.8$ & $2.5\times10^{41}$ & 2.5 & $6\times10^{-3}$ & 1.9 & 1.7 & $6\times10^{-3}$ & 3.5 & $<7$ \\
  W94 &3 & $L_{\rm bol}=10^{43}$ & $4\times10^{41}$ & 2.23 & $4\times10^{-2}$ & 3.86 & 2 & $6\times10^{-2}$ & 2 &~~~~~~$<9$
   \enddata
   \tablenotetext{}{A comparison of the analytic formula provided in this paper to numerical simulations presented in \cite{Bersten12} ($\Ren=270,200,$ and $150\,R_\odot$) and \cite{Woosley94} (model 13B). The numerical values include the relevant initial conditions and results of the simulations. The analytic values are calculated using equations (\ref{eq:Menv}),  (\ref{eq:Ren}), and  (\ref{eq:Rcor}) with initial conditions taken from the numerical simulations. The agreement of the numerical and analytical results is better than a factor of 2 (see discussion in the text).}
   \tablenotetext{a}{The extended material velocity is used in equations (\ref{eq:Menv}) and (\ref{eq:Ren}) when provided ($\ve=10^9\,{\rm cm\,s^{-1}}$ in model 13B of \citealt{Woosley94}). Otherwise, $E_{51}$ and $\Mc$ \citep[from][]{Bersten12} are used in these equations.}
   \tablenotetext{b}{The time of the first optical peak.}
   \tablenotetext{c}{The bolometric luminosity (from \citealt{Woosley94}) or the specific luminosity in the $g'$-band (from \citealt{Bersten12}) at the first peak.}
   \tablenotetext{d}{The pre-explosion mass within the radius range of $\Ren/3$ to $\Ren$ (see text for discussion).}
\label{Tbl:numericlComparison}
\end{deluxetable*}

Another property of the progenitor that can be constrained by the observations is $\Rc$. Since the emission from the core is similar to that of Type Ib/Ic SNe, the luminosity from the shock cooling phase decreases to a roughly constant minimal value before the $^{56}$Ni driven emission becomes dominant \citep{Dessart11}. In \cite{Piro13} we provide an analytic approximation for the minimal value of the shock cooling phase, and find that it is strongly correlated to the core radius (see their eq. [5]). Thus, observing this plateau provides a constraint on the core radius. However, during the first peak the emission is dominated by the extended material, while $^{56}$Ni decay, which dominates the second peak, already makes a significant contribution the minimum of the luminosity observed between the two peaks, $L_{\rm min}$. Therefore, the observed luminosity is always brighter than the minimal predicted value for the cooling shock emission of the core alone. Thus, $L_{\rm min}$ puts an upper limit on the core radius of
\be\label{eq:Rcor}
 	\Rc &\lesssim& 2.5 \times 10^{11}	\kappa_{0.2}^{0.9}
	\E51^{-1.1}
	\nonumber\\
	&&\times
	\lp \frac{L_{\rm min}}{10^{41}\,{\rm erg\,s^{-1}}}\rp^{1.3}
	 \lp \frac{\Mc}{3\Ms}\rp^{0.85}
	 {\rm cm},
\ee
where we use a canonical value of $\kappa_{0.2}=\kappa/0.2\,{\rm cm^2\,g^{-1}}$, as appropriate for a hydrogen deficient ionized gas. The connection between $L_{\rm min}$ and $R_{\rm core}$ is shown in red in Figure \ref{Fig2}. Note that the temperature during the rising phase of the second peak is typically in the optical and therefore $L_{\rm min}$ can be often estimated based on optical observations alone of the minimum between the two peaks.


\subsection{Comparison to Numerical Work}

In order to evaluate the accuracy of our analytic approximations we compare them to the results of detailed numerical simulations.  Explosion simulations of progenitors with low-mass, extended envelopes were carried out for two of the best studied SNe with double-peaked light curves of the type we consider here, 2011dh \citep{Bersten12} and 1993J \citep{Woosley94}. We compare our results to three models of \cite{Bersten12}, all of which have the same core structure and the same explosion energy ($E_{51}=1$), but the envelopes are extended to different radii of $\Ren=270,200,$ and $150\,R_\odot$. We also compare to model 13B of \cite{Woosley94}, which is found to produce a light curve that is similar to SN 1993J.

In all these simulations, a mass $\approx 0.1\Ms$ that contains hydrogen is attached to a $\approx 4\Ms$ He core. Most of this mass is concentrated right near the outer edge of the core radius, while a smaller amount of mass is spread over the extended parts of the envelope, around $\Ren$. As discussed above, due to adiabatic loses, the first peak is dominated by the emission from the mass near $\Ren$ and therefore, we take $\Me$ to be the mass between the radii of $\Ren/3$ and $\Ren$ right before the explosion.

A comparison between the numerical results and our formulas is presented in Table \ref{Tbl:numericlComparison}.  Our estimates  of $\Me$  and $\Ren$  agree very well, better than a factor of 2,  in all cases. We expected such agreement for $\Me$ and  the estimates of $\Ren$  for the case studied  by \cite{Woosley94}, where $L_{\rm bol}$  is given. The agreement,  however, with the values  of $\Ren$ estimated  for the  three cases  studied by  \cite{Bersten12}, where  only the absolute $g'$-band  magnitude is given, are better than  expected. It is probably not representative of  the true uncertainty in $\Ren$ in that case, which  is accurate only to within  an order of  magnitude when only  optical photometry is  known (see discussion below equation \ref{eq:Tpeak}). Finally,  the upper limits on $\Rc$ are all  a factor of $3-5$  larger than the actual  core radius.

\section{Summary}\label{Sec:Summary}
We have explored what can be learned about the progenitor properties from the light curves of SNe that show two peaks, where the first peak is seen also in the $R$ and/or $I$ bands and the second peak is powered by radioactive decay. We consider the emission from two types  of progenitors. Our main results are as follows.


{\it Standard progenitors.}
The planar  phase of  an extended (e.g., red supergiant) progenitor produces an optical peak with a rise time of $R_*/c\sim\,{\rm minutes}$ and a decay time of $R_*/v\sim\,{\rm hours}$. This phenomenon has yet to be seen in observations, but would be an important test of SN theory. The first optical peak in all known double-peaked SNe occur on a longer time scale and are not explained by this planar emission.

During the spherical phase, for both compact and extended standard progenitors, we derive an upper limit to how quickly the bolometric luminosity can drop. This is found to be $\alpha<0.64$,  where  $L_{\rm bol}\propto  t^{-\alpha}$.  For an  envelope structure with a typical polytropic index, the  limit is more  stringent, $\alpha < 0.35$. Furthermore,  the $R$  and $I$ band fluxes do not show a  significant decay (never faster than $L_{\rm bol}$) at any time because the temperature is either too high, or the gas recombines and its ionization level drops. These factors prevent standard progenitors from being able to produce the first peak in the $R$  and $I$ bands (note that it can produce two peaks in the UV and the blue optical bands, see Fig. \ref{Fig1}).

{\it Non-standard progenitors.}
We show that progenitors with extended, low-mass material on top of a compact, massive core naturally produce SNe with double-peaked light curves of the type we consider. The first peak is dominated by the cooling of the shock-heated extended material, and the second peak is the radioactive decay of $^{56}$Ni in the core. We show that the following properties can be constrained.
\begin{enumerate}
\item The time of the first peak provides a constraint on the extended material mass $\Me$ (eq. [\ref{eq:Menv}]).
\item The bolometric luminosity at $t_p$ measures the initial radius of the extended material $\Ren$ (eq. [\ref{eq:Ren}]). If only the specific luminosity at one or more optical bands is known, then $\Ren$ can still be constrained (using eq. [\ref{eq:Tpeak}] in addition), although less accurately.
\item The minimal observed luminosity, between the two peaks, sets an upper limit to the core radius $\Rc$ (eq. [\ref{eq:Rcor}]).
\end{enumerate}
Note that the time of the minimum between the two peaks is dominated by the decay rate of the first peak, and thus by $\Me$ and not by $\Ren$. This is consistent, for example, with the result of \cite{VanDyk13}, who find that the radius of the progenitor of SN 1993J is comparable to, or larger than, that of SN 2013df, even though the emission following the first peak of the latter decays more slowly.

The observed signatures that we discuss here are insensitive to the exact density profile of the extended material. The point where the details of the structure affect the light curve is the rise to the first peak. When the extended material is in hydrostatic equilibrium, then the light curve before the first peak is expected to follow the planar and spherical phases of a standard progenitor that we discussed here. Thus, very early future observations of double-peaked SNe have the potential to detect a {\it third peak} on the time scale of $\sim10\,{\rm min}$ after the explosion.

Finally, in this paper we focused on the question of what can be learned about the progenitor's density structure, but we ignored the problem of how stellar evolution can lead to such a progenitor. In the case of Type IIb SNe, it is generally thought that an interacting binary is responsible \citep[e.g.,][]{Podsiadlowski93,Stancliffe09,Eldridge2008,Claeys2011,Smith2011}. This has been confirmed most recently by binary models of SN 2011dh by \citet{Benvenuto2013}. However, it is not clear a hydrogen shell with a mass of $\approx10^{-3}-10^{-2}\,M_\odot$ should generically be expected by such mass transfer. If too little hydrogen is left at core collapse, then the first peak will not be present. In addition, mass transfer models by \citet{Yoon2010} find relatively compact progenitors, which would lead to a dim or non-existent first peak. An important question for future binary evolution studies is to understand which situations are best suited for a double-peaked light curve, and how often they should be expected.

The H deficient progenitors (e.g., SN 2006aj, iPTFbeo) are more difficult to understand, since there is currently no obvious stellar evolution model that leads to an explosion with the necessary structure. Some models find that a massive WR with a strong luminosity may inflate a small amount of mass to produce a core-halo structure \citep{Ishii99,Petrovic06,Grafener12}. However, the amount of inflated mass is too low ($\lesssim 10^{-6}\Ms$) to affect the light curve for more than $\sim 10\,{\rm min}$. An alternative option is a massive mass-loss episode that takes place just prior to the explosion. Occurring close enough to the explosion implies that it is likely causally connected to the final stages of evolution of the star. Recent observations suggest that late mass-loss episodes indeed take place \citep{Ofek13,Ofek14,Svirski14,Gal-yam14}. There are also recent theoretical models that predict increased mass loss prior to SN explosions \citep{Chevalier12,Shiode14}. In the coming years, a growing number of early SN light curves will provide important information about the structure of SN progenitors as they explode and on the evolution that brought them to these structures.


\acknowledgements
We thank A. Gal-Yam, D. Maoz, E. Ofek, C. Ott, and D. Poznanski for helpful comments. EN was partially supported by an ERC starting grant (GRB-SN 279369) and by the I-CORE Program of the
Planning and Budgeting Committee and The Israel Science Foundation (1829/12). ALP is supported through NSF grants AST-1205732, PHY-1068881, PHY-1151197, and the Sherman Fairchild Foundation

\bibliographystyle{apj}
\bibliography{msV2}
\end{document}